\documentclass{emulateapj}
\usepackage{apjfonts}
\usepackage{natbib}
\usepackage{epsfig}

\newcommand{\tq}{$t_{Q}$}
\newcommand{\tQ}{t_{Q}}
\newcommand{\etal}{et al.}
\newcommand{\NH}{$N_{\rm H}$}
\newcommand{\NHI}{$N_{\rm H\,I}$}
\newcommand{\nh}{N_{\rm H}}
\newcommand{\nhi}{N_{\rm H\,I}}
\newcommand{\Mdot}{\dot{M}}
\newcommand{\Lbol}{L_{\rm bol}}
\newcommand{\dEdt}{\epsilon_r \Mdot c^{2}}
\newcommand{\LB}{L_{B}}

\newcommand{\LBo}{L_{B,{\rm obs}}}
\newcommand{\LBm}{L_{B,{\rm min}}}

\newcommand{\Lcut}[1]{10^{#1}\,L_{\sun}}
\newcommand{\ctH}{\citet{Hopkins05}}

\shorttitle{Quasar Evolution}
\shortauthors{Hopkins \etal}
\slugcomment{ApJ, Accepted, September 2005}
\begin{document}

\title{Black Holes in Galaxy Mergers: Evolution of Quasars}
\author{Philip F. Hopkins\altaffilmark{1}, 
Lars Hernquist\altaffilmark{1}, 
Thomas J. Cox\altaffilmark{1}, 
Tiziana Di Matteo\altaffilmark{2}, 
Paul Martini\altaffilmark{1}, 
Brant Robertson\altaffilmark{1}, 
Volker Springel\altaffilmark{3}}
\altaffiltext{1}{Harvard-Smithsonian Center for Astrophysics, 
60 Garden Street, Cambridge, MA 02138, USA}
\altaffiltext{2}{Carnegie Mellon University, 
Department of Physics, 5000 Forbes Ave., Pittsburgh, PA 15213}
\altaffiltext{3}{Max-Planck-Institut f\"{u}r Astrophysik, 
Karl-Schwarzchild-Stra\ss e 1, 85740 Garching bei M\"{u}nchen, Germany}

\begin{abstract}

Based on numerical simulations of gas-rich galaxy mergers, we discuss
a model in which quasar activity is tied to the self-regulated growth
of supermassive black holes in galaxies.  The nuclear inflow of gas
attending a galaxy collision triggers a starburst and feeds black hole
growth, but for most of the duration of the starburst, the black hole
is ``buried'', being heavily obscured by surrounding gas and dust,
limiting the visibility of the quasar, especially at optical and
ultraviolet wavelengths.  As the black hole grows, feedback energy
from accretion heats the gas and eventually expels it in a powerful
wind, leaving behind a ``dead quasar''.  In between the buried and
dead phases, there is a window in time during which the galaxy would
be seen as a luminous quasar.  Because the black hole mass, radiative
output, and distribution of obscuring gas and dust all evolve strongly
with time, the duration of this phase of observable quasar activity
depends on both the waveband and imposed luminosity threshold.  We
determine the observed and intrinsic lifetimes as a function of
luminosity and frequency, and calculate observable lifetimes
$\sim10\,$Myr for bright quasars in the optical B-band, in good
agreement with empirical estimates and much smaller than our estimated
black hole growth timescales $\sim100\,$Myr, naturally producing a
substantial population of buried quasars.  However, the observed and
intrinsic energy outputs converge in the IR and hard X-ray bands as
attenuation becomes weaker and chances of observation greatly
increase.  We also obtain the distribution of column densities along
sightlines in which the quasar is seen above a given luminosity, and
find that our result agrees remarkably well with observed estimates of
the column density distribution from the SDSS for the appropriate
luminosity thresholds.  Our model reproduces a wide range of quasar
phenomena, including observed quasar lifetimes, intrinsic lifetimes,
column density distributions, and differences between optical and
X-ray samples, having properties consistent with observations across
more than five orders of magnitude in bolometric luminosity from
$10^{9}-10^{14}\,L_{\sun}$ ($-17\lesssim M_{B}\lesssim -30$).

\end{abstract}


\section{Introduction\label{sec:intro}}

More than 40 years have passed since the recognition that quasars are
at cosmological distances (Schmidt 1963; Greenstein \& Mathews 1963;
Mathews \& Sandage 1963) and hence must be powerful energy sources.  A
combination of arguments based on time variability and energetics
strongly supports the view that the activity is produced by the
accretion of gas onto supermassive black holes in the centers of
galaxies (e.g. Salpeter 1964; Zel'dovich \& Novikov 1964; Lynden-Bell
1969).  However, the mechanism that provides the trigger to fuel
quasars remains uncertain.  Recent discoveries of correlations between
masses of black holes in nearby galaxies and either the mass
(Magorrian et al. 1998) or velocity dispersion (i.e. the $M_{\rm
BH}$-$\sigma$ relation: Ferrarese \& Merritt 2000; Gebhardt et
al. 2000) of spheroids demonstrate a fundamental link between the
growth of supermassive black holes and galaxy formation.  However,
theoretical evidence connecting the origin of supermassive black holes
and quasars to galaxy evolution remains elusive, owing to the
complexity of the underlying physics and dynamics.  In particular, up
to now there has been no comprehensive model to explain the origin and
fueling of quasars, or their lifetimes, obscuration, demographics,
self-regulation and termination, and dependence on host galaxy
properties.

Observations of the nearby Universe suggest that rapid black hole
growth may be related to massive flows of gas into the centers of
galaxies.  Infrared (IR) luminous galaxies represent the dominant
population of objects above $10^{11} L_\odot$ locally and it is
believed that much of their IR emission is powered by dust
reprocessing of radiation from intense nuclear starbursts (e.g. Soifer
et al. 1984a,b; Sanders et al. 1986, 1988a,b; for a review, see
e.g. Soifer et al. 1987).  At the highest luminosities above $10^{12}
L_\odot$, characteristic of ultraluminous infrared galaxies (ULIRGs),
nearly all the galaxies appear to be in advanced stages of merging
(e.g. Allen et al. 1985; Joseph \& Wright 1985; Armus et al. 1987;
Kleinmann et al. 1988; for reviews, see Sanders \& Mirabel 1996 and
Jogee 2004) and CO observations show that they contain large
quantities of gas in their nuclei (e.g. Scoville et al. 1986; Sargent
et al. 1987, 1989).  Some ULIRGs exhibit ``warm'' IR spectral energy
distributions, perhaps indicative of a buried quasar (e.g. Sanders et
al. 1988c).  This fact, together with the overlap between bolometric
luminosities of ULIRGs and quasars, led Sanders et al. (1988a) to
propose that quasars are the descendents of an infrared luminous phase
of galaxy evolution caused by mergers.  This scenario is supported by
recent X-ray observations which have revealed the presence of two
non-thermal point sources near the center of the ULIRG NGC6240
\citep{Komossa03}, which are most naturally interpreted as accreting
supermassive black holes that are heavily obscured at visual
wavelengths (e.g.  Gerssen et al. 2004; Max et al. 2005).

Hydrodynamic simulations have shown that gas inflows are produced by
strong gravitational torques on the gas through tidal forces during
mergers involving gas-rich galaxies
\citep[e.g.,][]{Hernquist89,BH91,BH96}.  The resulting dense
concentrations of gas in the inner regions of the remnant can sustain
star formation at rates high enough and for sufficiently long
timescales to account for many properties of ULIRGs
\citep[e.g.,][]{MH94,MH96,HM95}.  However, these authors were not
able to directly explore the relationship between the gas inflows
and quasar activity because their simulations did not include a
model for black hole growth and the impact of strong feedback from
either star formation or quasars.

One of the most fundamental parameters of black hole growth is the
quasar lifetime, \tq, which sets the timescale for the most luminous
phase of the activity.
Observations generally constrain quasar lifetimes to
the range $\tQ\approx 10^{6}-10^{8}$ yr \citep[for a review,
see][]{Martini04}.  These estimates are primarily based on demographic
or integral arguments which combine observations of the present-day
population of supermassive black holes and accretion by the
high-redshift quasar population
\citep[e.g.,][]{Soltan82,HNR98,YT02,YL04,HCO04}, or incorporate quasars into
models of galaxy evolution \citep[e.g.,][]{KH00,WL02,DCSH03,DCSH04,Gran04}
or reionization of HeII (e.g. Sokasian et al. 2002, 2003).
Results from clustering in quasar surveys
\citep[e.g.,][]{PMN04,Grazian04}, the proximity effect in the
Ly$\alpha$ forest (Bajtlik, Duncan \& Ostriker 1988; Haiman \&
Cen 2002; Yu \& Lu 2005; although
such results may still be inconclusive, 
e.g. Croft 2004), and the transverse
proximity effect in He II \citep{Jakobsen03} similarly suggest
lifetimes $\tQ\sim10^{7}\,$yr.

It is not immediately obvious how quasar lifetimes $\tQ\sim10^{7}\,$yr
can be explained in the context of gas inflows triggered by galaxy
mergers.  The full duration of the starburst phase is set by the 
timescale
during which strong gas inflows are excited, which in turn is
determined by the time when significant gravitational torques are exerted
on the gas.  The numerical simulations of \citet{MH94,MH96} showed
that this occurs for $\sim 2\times 10^{8}$ yr, much shorter than
typical merger timescales of $> 10^{9}\,{\rm yr}$
\citep{B88,B92,H92,H93}, but in good agreement with observed estimates
for the gas depletion time in the central regions of ULIRGs (for a
discussion, see e.g. Barnes \& Hernquist 1992).  If gas-rich mergers
are indeed responsible for the origin of quasar activity, as suggested
through studies of ULIRGs (e.g. Sanders et al. 1988a) and more
directly from observations of quasar hosts (e.g.  Stockton 1978;
Heckman et al. 1984; Stockton \& MacKenty 1987; Stockton \&
Ridgway 1991;
Hutchings \& Neff 1992;
Bahcall et al. 1994, 1995;
Canalizo \& Stockton 2001), the factor $\sim 10$
difference between the gas consumption time in ULIRGs and quasar
lifetimes must be reconciled.

Semi-analytical models of supermassive black hole evolution and its
correlation with galaxy structure suggest that, beyond a certain
threshold, feedback energy expels nearby gas and shuts down the
accretion phase \citep{SR98,Fabian99,WL03}.  However, these
calculations provide only limits, and in neglecting the
dynamics of quasar evolution they do not predict time-dependent
effects such as the characteristic lifetime of the accretion phase
prior to its self-termination, the fueling rates for black hole
accretion, the obscuration of central sources, or the quasar light
curve. These quantities are instead usually taken to be independent
input parameters of the model. For example, the lifetime is either
adopted from observational estimates or assumed to be similar to a
characteristic timescale such as the dynamical time of the host galaxy
disk or the $e$-folding time for Eddington-limited black hole growth
$t_{S}=M_{\rm BH}/\Mdot=4.5\times10^{7}\,l\,(\epsilon_r/0.1)\,$yr for accretion
with radiative efficiency $\epsilon_r=L/\Mdot c^{2}\sim0.1$ and
$l=L/L_{Edd}\lesssim1$ \citep{Salpeter64}.

Efforts to model quasar accretion and feedback in a more self-consistent manner 
\citep[e.g.,][]{CO97,CO01} have considered the time-dependent 
nature of gas inflows feeding accretion and the impact of radiative heating 
from quasar feedback in a hydrodynamical context. Although generally 
restricted to one-dimensional ``toy models'' which do not include galaxy-galaxy 
interactions, such modeling has made important progress in predicting 
the characteristic duty cycles of quasars, in good agreement with 
e.g., \citet{HCO04}, 
and in demonstrating the importance of radiative feedback on the surrounding medium
and quasar evolution \citep[for a review, see][]{OC05}. 
Further, \citet{Sazonov05} showed that self regulation by 
radiative feedback from quasars in this modeling 
can expel a significant quantity of 
gas and leave a remnant with properties similar to observed 
ellipticals and black hole mass corresponding to the observed 
$M_{\rm BH}$-$\sigma$ relation. Similarly \citet{Kawata05} showed
that incorporating a simple prescription for radiative heating by quasars 
in the context of cosmological simulations yields populations of elliptical 
galaxies consistent with observed optical and X-ray properties.  

Recently, \citet{SDH05b} have developed a methodology for
incorporating black hole growth and feedback into hydrodynamical
simulations of galaxy mergers that includes a multiphase model for star
formation and pressurization of the interstellar gas by supernova
feedback \citep{SH03}.  Using this approach, we have begun to explore
the impact of these processes on galaxy formation and evolution.  Di
Matteo et al. (2005) and Springel et al. (2005b) have shown that the
gas inflows produced by gravitational torques during a merger both
trigger starbursts (as in the earlier simulations of e.g.  Mihos \&
Hernquist 1994, 1996) and fuel rapid black hole growth.  The growth of
the black hole is determined by the gas supply and terminates abruptly
when significant gas is expelled owing to the coupling between
feedback energy from black hole accretion and the
surrounding gas.  Eventually, as the gas is heated and driven out, the
remnant is no longer active because the black hole does not accrete at
a high rate, leaving a dead quasar in an ordinary galaxy.  The
self-regulated nature of the black hole growth explains observed
correlations between black hole mass and properties of normal galaxies
\citep{DSH05}, as well as the color distribution of ellipticals
\citep{SDH05a}.  These results lend support to the view that mergers
have played an important role in structuring galaxies, as advocated by
e.g. Toomre \& Toomre (1972) and Toomre (1977).

Moreover, \citet{DSH05} showed that the dynamics of the inflowing gas
and its response to the self-regulated growth of the black hole yield
a timescale for the strong accretion phase $\sim 2\times 10^{8}$ yr,
comparable to the full duration of the starburst.  During much of this
period, the bolometric luminosity of the black hole would exceed the
threshold to be classified as a quasar, implying a timescale $\sim
10^{8}\,{\rm yr}$ (for a Milky Way mass system) for the intrinsic
quasar phase of the black hole.

In \ctH, we employed models for obscuration by gas and dust to show
that for most of the accretion lifetime the quasars would be buried;
i.e. heavily obscured by the large gas density powering accretion.
Eventually, feedback from the accretion energy drives away the gas,
creating a brief window in which the central object would be observable
as an optical quasar, until accretion levels drop below quasar
thresholds.  By calculating the effects of obscuration from the
simulation of a merger of gas-rich galaxies, we showed that this
determines an observable lifetime $\sim 10^{7}\,{\rm yr}$, in good
agreement with observations.

Here, we extend and further develop implications of the model for
quasar evolution proposed in \ctH\ by analyzing a series of
hydrodynamical simulations of galaxy mergers where we vary 
the masses of the progenitor galaxies, so that they have
virial velocities between 80 and $320\,{\rm km\,s^{-1}}$.  In
\S2\ %\ref{sec:modelcalc} 
we describe our series of merger
simulations and the calculation of column densities and obscuration,
and compare the results obtained with different column density
calculations. In \S3\ %\ref{sec:freq} 
we examine the frequency dependence
of our results and quantify the differences in lifetimes across the
optical and X-ray bands.  In \S4\ %\ref{sec:NHdistrib} 
we examine the typical column densities along lines-of-sight to the simulated
quasars, and compare these to observed distributions from both optical
and X-ray surveys.  In \S5\ %\ref{sec:lifetimes} 
we show the simulation
results for our merger series and compare the observed and intrinsic
lifetimes as a function of luminosity for all the
simulations. Finally, in \S6\ %\ref{sec:discussion} 
we discuss our results
and the implications of our model for a broad range of quasar studies.

\section{The Model: Calculations\label{sec:modelcalc}}
\subsection{The Simulations\label{sec:sim}}

The simulations were performed using the GADGET-2 code, a new version
of the parallel TreeSPH code GADGET \citep{SYW01}. It uses an
entropy-conserving formulation \citep{SH02} of smoothed particle
hydrodynamics (SPH), and includes radiative cooling, heating by a UV
background (as described in e.g. Katz et al. 1996; Dav\'e et al. 1999),
and a sub-resolution model of a multiphase interstellar
medium (ISM) to describe star formation and supernova feedback
\citep{SH03}. This sub-resolution model provides an effective equation
of state for star-forming gas which includes pressure feedback from
supernova heating, and allows us to stably evolve even massive pure
gas disks (see, e.g. Springel et al. 2005b;
Robertson et al. 2004). The methodology of
accretion, feedback, and galaxy generation is described in detail in
\citet{SDH05b}.

In our approach, supermassive black holes (BHs) are represented by
``sink'' particles that accrete gas from their local environment, with
an accretion rate $\Mdot$ estimated from the local gas density and
sound speed using a Bondi-Hoyle-Lyttleton parameterization with an
imposed upper limit equal to the Eddington rate. The bolometric
luminosity of the BH particle is then $\Lbol=\dEdt$, where
$\epsilon_r=0.1$ is the radiative efficiency. We further assume that a
small fraction ($5\%$) of $\Lbol$ couples dynamically to the
surrounding gas, and that this feedback is injected into the gas as
thermal energy. This fraction is a free parameter, determined in
\citet{DSH05} by fitting to the $M_{\rm BH}$-$\sigma$ relation.  We do
not attempt to resolve the small-scale accretion dynamics near the
black hole, but instead assume that the time-averaged accretion can be
estimated from the gas properties on the scale of our spatial
resolution ($\lesssim 50$\,pc).

\begin{deluxetable}{cccc}
\tabletypesize{\scriptsize}
\tablecaption{Simulation Parameters\label{tbl:sims}}
\tablewidth{0pt}
\tablehead{
\colhead{Simulation} & \colhead{$V_{\rm vir}\,({\rm km\,s^{-1}})$} & 
\colhead{$M_{\rm vir}\,({\rm M_{\sun}})$} & 
\colhead{$M_{\rm BH,\, final}\,({\rm M_{\sun}})$}
}
\startdata
A1 & 80 & $1.7\times10^{11}$ & $7\times10^{6}$ \\
A2 & 113 & $4.8\times10^{11}$ & $3\times10^{7}$ \\
A3 & 160 & $1.4\times10^{12}$ & $3\times10^{8}$ \\
A4 & 226 & $3.8\times10^{12}$ & $7\times10^{8}$ \\
A5 & 320 & $1.1\times10^{13}$ & $2\times10^{9}$ \\
\enddata
\end{deluxetable}

In what follows, we analyze five simulations of colliding disk
galaxies, which form a family of structurally similar models with different
virial velocity and mass.  In each simulation, we generate two stable,
isolated disk galaxies, each with an extended dark matter halo with a
\citet{Hernquist90} profile, motivated by cosmological
simulations (e.g. Navarro et al. 1996; Busha et al. 2004) and
observations of halo properties (e.g. Rines et al. 2002, 2002, 2003, 2004),
an exponential gas disk, and a bulge. The
simulations follow the series described in detail in \citet{SDH05a},
with the parameters listed in Table~\ref{tbl:sims}. We denote the
simulations A1, A2, A3, A4, and A5, with increasing virial velocities
of $V_{\rm vir}=80, 113, 160, 226, {\rm and}\ 320\,{\rm km\,s^{-1}}$,
respectively.  Note that the self-similarity of these models is broken
by the scale-dependent physics of cooling, star formation, and black
hole accretion.  In \ctH, we describe our analysis of simulation A3, a
fiducial choice with a rotation curve and mass similar to the Milky
Way.  The galaxies have mass $M_{\rm vir}=V_{\rm vir}^{3}/(10GH_{0})$,
with the baryonic disk having mass fraction $m_{\rm d}=0.041$, the
bulge $m_{\rm b}=0.0136$, and the rest of the mass in dark matter with
a concentration parameter $9.0$.  The disk-scale length is computed
based on an assumed spin parameter $\lambda=0.033$, and the
scale-length of the bulge is set to $0.2$ times this.  We begin our
simulation with pure gas disks, which may better correspond to the
high-redshift galaxies in which most quasars are observed.

Each galaxy is initially composed of 168000 dark matter halo
particles, 8000 bulge particles, 24000 gaseous disk particles, and one
BH particle, with a small initial seed mass of $10^{5}M_{\sun}$. Given
these choices, the dark matter, gas, and star particles are all of
roughly equal mass, and central cusps in the dark matter and bulge
profiles are reasonably well resolved (see Fig
2. in Springel et al. 2005b).  The galaxies are then set to collide in
a prograde encounter with zero orbital energy and a small pericenter
separation ($7.1\,{\rm kpc}$ for the A3 simulation).

\subsection{Column Densities \&\ Quasar Attenuation\label{sec:NH}}

We calculate the column density between a black hole and a
hypothetical observer from the simulation outputs spaced every 10 Myr
before and after the merger and every 5 Myr during the merger of each
galaxy pair. The calculation method is described in \ctH, but we review it
briefly here. We generate $\sim1000$ radial lines-of-sight (rays),
each with its origin at the black hole particle location and with
directions uniformly spaced in solid angle $d\cos{\theta}\,d\phi$. For
each ray, we then begin at the origin, calculate and record the local
gas properties using GADGET, and then move a distance along the ray
$\Delta r=\eta h_{\rm sml}$, where $\eta \leq 1$ and $h_{\rm sml}$ is
the local SPH smoothing length. The process is repeated until a ray is
sufficiently far from its origin ($\gtrsim 100$ kpc). The gas
properties along a given ray can then be integrated to give the
line-of-sight column density and mean metallicity.  We test different
values of $\eta$ and find that gas properties along a ray converge
rapidly and change smoothly for $\eta=0.5$ and smaller. We similarly
test different numbers of rays and find that the distribution of
line-of-sight properties converges for $\gtrsim 100$ rays.

Given the local gas properties, we use the GADGET multiphase model of
the ISM described in \citet{SH03} to calculate the local mass fraction
in ``hot'' (diffuse) and ``cold'' (molecular and HI cloud core) phases
of dense gas and, assuming pressure equilibrium between the two phases,
we obtain the local density of the hot and cold phase gas and the
corresponding volume filling factors. The values obtained correspond roughly
to the fiducial values of \citet{MO77}.  Given a temperature for the
warm, partially ionized component $\sim8000\,{\rm K}$, determined by
pressure equilibrium, we further calculate the neutral fraction of
this gas, typically $\sim0.3-0.5$.  We denote the neutral and total
column densities as \NHI\ and \NH, respectively. Using only the
hot-phase density allows us to place an effective lower limit on the
column density along a particular line of sight, as it assumes a ray
passes only through the diffuse ISM, with $\gtrsim 90\%$ of the mass
of the dense ISM concentrated in cold-phase ``clumps.'' Given the
small volume filling factor ($<0.01$) and cross section of such
clouds, we expect that the majority of sightlines will pass only
through the ``hot-phase'' component.

Using $\Lbol=\dEdt$, we model the form of the intrinsic quasar
continuum SED following \citet{Marconi04}, based on optical through
hard X-ray observations
\citep[e.g.,][]{Elvis94,George98,VB01,Perola02,Telfer02,Ueda03,VBS03}.
This gives a B-band luminosity
$\log{(\LB)}=0.80-0.067\mathcal{L}+0.017\mathcal{L}^{2}-0.0023\mathcal{L}^{3}$,
where $\mathcal{L} = \log{(\Lbol/L_{\sun})} - 12$, and we take
$\lambda_{B}=4400\,$\AA.  We then use a gas-to-dust ratio to determine
the extinction along a given line of sight at this frequency.
Observations suggest that the majority of the population of reddened
quasars have reddening curves similar to that of the Small Magellenic
Cloud (SMC; Hopkins et al. 2004), which has a gas-to-dust ratio lower
than the Milky Way by approximately the same factor as its metallicity
\citep{Bouchet85}. We therefore consider both a gas-to-dust ratio
equal to that of the Milky Way, $(A_{B}/\nhi)_{\rm
MW}=8.47\times10^{-22}\,{\rm cm^{2}}$, and a gas-to-dust ratio scaled
by metallicity, $A_{B}/\nhi = (Z/0.02)(A_{B}/\nhi)_{\rm MW}$. For both
cases we use the SMC-like reddening curve of \citet{Pei92}. We
calculate extinction in X-ray frequencies (0.03-10 keV) using the
photoelectric absorption cross sections of \citet{MM83} and
non-relativistic Compton scattering cross sections, similarly scaled
by metallicity. In determining the column density for photoelectric
X-ray absorption, we ignore the inferred ionized fraction of the
gas, as it is expected that the inner-shell electrons which dominate
the photoelectric absorption edges will be unaffected in the
temperature ranges of interest.  We do not perform a full radiative
transfer calculation, and therefore do not model scattering or
re-processing of radiation by dust in the infrared.

\subsection{Quasar Lifetimes \&\ Sensitivity to Simulation Parameters\label{sec:detailsCompare}}

We quantify the quasar lifetime \tq\ as in \ctH, as a function of the
limiting B-band luminosity $\LBm$. For each sightline, an observed
lifetime is determined as the integrated time in the simulation during
which the given sightline sees a B-band luminosity above a given
threshold, $\LBo>\LBm$. We consider both the intrinsic lifetime 
$\tQ^{\,i}$, the total time the
intrinsic $\LB\geq \LBm$ (ignoring attenuation), and the observed lifetime 
derived from a particular column density calculation.
We note that below 5 Myr
(the simulation output frequency in the standard case) our estimates
of \tq\ become uncertain owing to the effects of quasar variability
and our inability to resolve the local small-scale physics of the
ISM. However, the majority of sightlines see lifetimes in the range
$10-20\,$Myr up to $\LBm=\Lcut{11}$, in good agreement with
observations suggesting lifetimes $\sim10^{7}\,$yr and well above this
limit.

\begin{figure}
    \centering
    %\plotone{f1.ps}
    \includegraphics[width=3.4in]{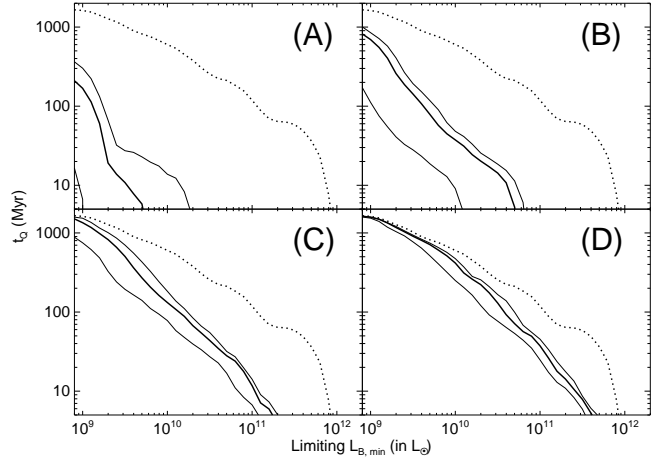}
    \caption{Quasar lifetimes as a function of $\LBm$ from the A3
    simulation.  The dotted lines show the lifetime if attenuation is
    ignored (time with observed B-band luminosity $\LBo\geq
    \LBm$). The thick solid lines indicate the median observed lifetime,
    with thin lines giving the 25\%-75\% contours.  Lifetimes are
    calculated using: (A) cold-phase ISM density, (B) hot-phase
    density (ignoring metallicity and ionization), (C) hot-phase
    density (including metallicity and ionization), (D) hot-phase
    density with artificially low metallicity (0.1 times solar).
    \label{fig:compareNHcalc}}
\end{figure}

Figure~\ref{fig:compareNHcalc} shows the lifetime obtained in the A3
simulation using different calculations of the column density \NHI\ (cases $A-D$). The
uppermost curve in all cases is the intrinsic lifetime
$\tQ^{\,i}$. The lower thick curve is the median observed lifetime
calculated using a particular estimate of the column density, with the
thin curves representing the 25\%-75\% inclusion contours. The
cold-phase ISM density ($A$) is large enough for the quasar to be
completely extincted
out of observable ranges for the duration of the quasar phase,
rendering the object observable only during quiescent phases with
$\LB\sim\Lcut{9}$. We also show the significantly longer lifetimes
calculated using the hot-phase ISM density, first ignoring corrections
for metallicity and ionization 
($B$; i.e.\ assuming $A_{B}/\nhi=(A_{B}/\nhi)_{\rm MW}$
and $\nhi=\nh$)
and then including these effects ($C$), which
increases the median lifetime by a factor $\sim2$ and significantly
reduces the scatter towards shorter lifetimes. Finally, we calculate
the lifetime using the hot-phase density and an SMC-like gas-to-dust
ratio, $(A_{B}/\nhi)_{\rm SMC}=0.869\times10^{-22}\,{\rm cm^{2}}$
($D$; essentially assuming a metallicity $0.1$ times solar), which sets a
strong upper limit on the observed lifetime. These lifetimes are still
only $\sim2$ times as long as the lifetimes calculated using
metallicity-weighted neutral column densities, and still $\sim1/3$ the
intrinsic lifetime at $\LBm=\Lcut{11}$.  We therefore expect that,
after accounting for the clumping of most mass in the cold phase of
the most dense regions of the ISM, 
the qualitative relation between our observed and intrinsic 
calculated lifetimes should be
relatively insensitive to the details of the column density
calculation, with variation by a factor of $\sim2-3$ between 
lifetimes calculated using different prescriptions for the multiphase 
ISM, ionization, and metallicity effects. 

\section{Frequency Dependence of the Quasar Lifetime\label{sec:freq}}

We expect the intrinsic quasar lifetime above a given luminosity in a
particular band to change with the frequency of that band. The shift in
lifetime with frequency is approximately given by an offset in
luminosity corresponding to the difference in luminosities at different
frequencies in our model quasar spectrum. However, this change is not
entirely history-independent, as the model spectrum alters shape with
varying luminosity owing to bolometric corrections. More
important, our model predicts that the observed lifetime above a
given luminosity will change with frequency not only as a result of
the shift in intrinsic lifetime, but primarily as a result of
varying levels of attenuation at different frequencies.

\begin{figure}
    \centering
    %\plotone{f2.ps}
    \includegraphics[width=3.5in]{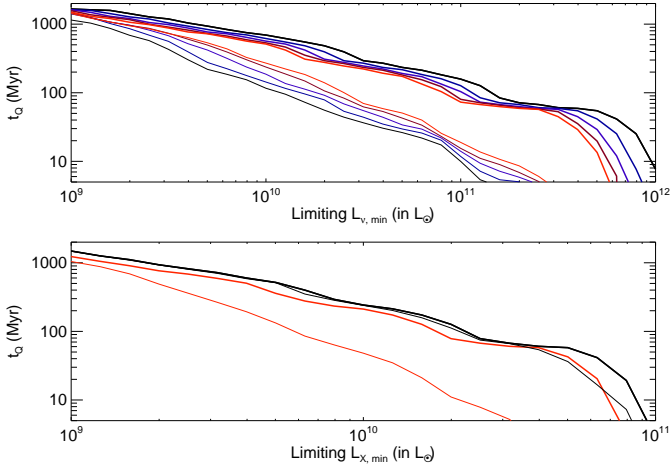}
    \caption{Quasar lifetimes as a function of $\LBm$ from the A3
    simulation, showing the waveband dependence of the lifetime.
    The thick lines give the lifetime if attenuation is
    ignored, thin lines indicate the median observed lifetime
    (calculated using the metallicity-weighted, hot-phase neutral column density). Upper
    panel: From black to blue to red, lifetimes are calculated at 3000, 4500,
    6000, 7500, and 9000 \AA. Lower panel: lifetimes are calculated
    for soft ($0.5-2\,{\rm keV}$, red) and hard ($2-10\,{\rm keV}$,
    black) X-ray bands.
    \label{fig:lifetimeVSfreq}}
\end{figure}

Figure~\ref{fig:lifetimeVSfreq} shows the lifetimes as a function of
luminosity for the A3 simulation, for various frequencies in the
optical and X-ray bands. We plot lifetimes in the manner of
Figure~\ref{fig:compareNHcalc}; i.e. the total time with $\nu
L_{\nu}>L_{\nu, min}$ ($L_{\nu, min}\equiv (\nu dL/d\nu)_{min}$), 
at the representative optical wavelengths
$\lambda=3000, 4500, 6000, 7500, {\rm and}\ 9000\,$\AA\ (colored black
to blue to red, respectively). We find that the intrinsic lifetime
systematically decreases at longer wavelengths (based on the shape of
the quasar SED), while the observed lifetime increases (owing to
weakening attenuation), with the two approaching one another near
infrared wavelengths. At longer wavelengths, we expect the curves to
cross, as the observed lifetime in the far IR, owing to dust
re-processing of quasar radiation, will be longer than the intrinsic
lifetime, but we do not yet incorporate such re-processing into our
column density calculations. At very low $L_{\nu, min}$, lifetimes 
approximately converge to the total duration of the simulations, as  
the quasar is above such luminosities throughout almost the entire merger. 

The lower panel of Figure \ref{fig:lifetimeVSfreq} shows the lifetimes
in the soft and hard X-ray bands, defined as $0.5-2\,{\rm keV}$ and
$2-10\,{\rm keV}$, respectively. We find that the soft X-ray band is
heavily attenuated and has much shorter observed than intrinsic
lifetimes, primarily owing to photoelectric absorption. However, the
hard X-ray band is relatively unaffected, as the hot-phase column
densities tend to be well below the Compton-thick regime
$\nh\sim10^{24}\,{\rm cm^{-2}}$. Thus, most quasars obscured in the
optical for much of their high accretion rate lifetimes should be
observable in hard X-rays. We defer a full treatment of the difference
between the attenuation of soft and hard X-rays to a future paper, but
note that this difference may account for the slope of the cosmic
X-ray background, with a population of obscured quasars at spectral
energies $\lesssim$ a few keV as a natural stage in the evolution of
quasar activity.

\section{Typical Column Densities\label{sec:NHdistrib}}

Given a calculation of the column density along multiple
lines-of-sight to the simulated quasars, we can examine the typical
column density distributions at all times and at times in which the
quasar is observed above some luminosity threshold. Following the
analysis in \ctH, we expect the characteristic  column densities to rise
rapidly during the merger, as gas inflow traces and fuels a rising
accretion rate. Eventually, column densities rapidly fall in a blowout
phase once the black hole has reached a critical mass, creating a
window with rapidly changing column densities in which the quasar is
observable until accretion rates drop below those necessary to fuel
quasar activity.  As is clear in Figure 2 of \ctH, the dispersion in
\NH\ at any particular time in the simulation is generally small, a
factor of $\sim2$ in either direction, but typical \NH\ values can
change by an order of magnitude over timescales
$\sim\,$Myr. Therefore, it is of interest to consider the probability
of an observer, viewing the object at a random time, seeing a given
column density.

\begin{figure}
    \centering
    %\plotone{f3.ps}
    \includegraphics[width=3.7in]{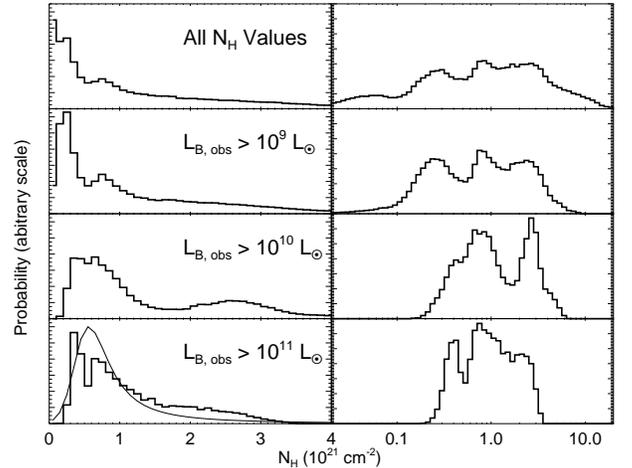}
    \caption{Distribution of ``hot-phase'', neutral column
    densities \NHI\ from the A3 simulation.  The probability of seeing
    a given \NHI\ is weighted by the time such an \NHI\ is observed
    along all sightlines. Plots are scaled linearly (left) and
    logarithmically (right). The top panel shows all \NHI\ values from
    all times, the lower panels give the distribution for times where
    the observed B-band luminosity is above the $\LBm$ shown. The
    smooth curve is the distribution of \NHI\ values from
    the observed SDSS quasar sample of 
    \citet{Hopkins04}, who used an absolute $i$-band magnitude limit
    approximately equivalent to $\LBo>\Lcut{11}$, scaled to the
    appropriate modal \NHI.
    \label{fig:NHdistrib}}
\end{figure}

Figure \ref{fig:NHdistrib} plots the distribution of column densities
\NHI\ calculated using the hot-phase metallicity-weighted ISM density
in the A3 simulation. The probability of seeing a given \NHI\ is
calculated as proportional to the total time along all sightlines that
such an \NHI\ is observed. The plots are scaled both linearly (left
panels) and logarithmically (right panels) in column density. Typical column densities
are distributed about $\nhi=10^{21}\,{\rm cm^{-2}}$, approximately
symmetrically in $\log{\nhi}$. We further plot the distribution of
\NHI\ values requiring that the observed B-band luminosity be above
some threshold $\LBm$. The curve shown in the linear plots is the
best-fit to the $E_{B-V}$ distribution of bright SDSS quasars with
$z<2.2$, from \citet{Hopkins04}. The curve has been rescaled in terms
of the column density (inverting our gas-to-dust prescription) and
plotted about a peak (mode) \NHI\ (undetermined in Hopkins et al. 2004) of
$\nhi\approx0.5\times10^{21}\,{\rm cm^{-2}}$. The $i$-band absolute
magnitude limit imposed in the observed sample, $M_{i}<-22$,
corresponds approximately to our plotted B-band limit
$\LBo>\Lcut{11}$.

The agreement between the observed column density distribution and the
result of our simulations once the same selection effect is applied is
strong evidence in favor of our model for quasar evolution. The distribution as a
function of limiting luminosity is a natural consequence of the
dynamics of the quasar activity. Throughout much of the duration of
bright quasar activity, column densities rise to high levels as a
result of the same process that feeds accretion, naturally producing a
reddened population of quasars (a red ``tail'' in the quasar color
distribution), extending to very bright quasars strongly reddened
by large \NHI. The existence of this extended reddened quasar
population in radio and optically selected quasars is well known
\citep[e.g.][]{Webster95,Brotherton01,Francis01,
Richards01,Gregg02,White03}, and we are able to reproduce its
distribution not as a distribution of source properties but as a
result of the time evolution of quasar phenomena.  Our estimate of the
distribution of \NHI\ for $\LBo>\Lcut{11}$ does not account for an
additional selection effect, namely that strongly reddened quasars may
not be extincted from an observed sample (if the intrinsic luminosity
is large enough), but their colors may be significantly reddened to
the point where color selection criteria of quasar surveys will not
include them. However, this effect would only serve to bring our
distribution into better agreement with observations, as it would
slightly lower the high-\NHI\ tail.

Moreover, our estimates of the distribution of \NHI\ allow us to
make a directly observable prediction, namely that the column density
distribution in optically selected samples with a minimum absolute
magnitude or luminosity should broaden in both directions (to larger
and smaller \NHI) as the limiting selection luminosity is decreased
(limiting absolute magnitude increased). This is because, at lower
luminosities, observers will see both intrinsically bright periods
extincted by larger column densities (broadening the distribution to
larger \NHI\ values) and intrinsically faint periods with small column
densities (broadening the distribution to smaller \NHI\ values). The
distribution of \NH\ values in hard X-ray quasar samples, which are
much less affected by this extinction (see \S3), should %\ref{sec:freq}
also reflect this trend, and indeed the distribution of \NH\ values is
much broader and flatter, with quasars seen from
$\nh\sim10^{20}-10^{24}\,{\rm cm^{-2}}$ \citep{Ueda03}.

Extension of this distribution to the other black hole masses and peak
quasar luminosities of our simulations reveals similar qualitative
behavior. The characteristic column densities during the obscured,
intrinsically bright phases of accretion generally increase with
galaxy and black hole mass, but typically by factors $\lesssim10$
across the range of our simulations $V_{\rm vir}=80-320\,{\rm
km\,s^{-1}}$. Additionally, although mean column densities increase
somewhat with increasing mass, the functional form of the \NHI\
distribution remains similar at each limiting luminosity, implying
that a combined population of quasars similar to those we have
simulated will match the observed \NHI\ distribution with appropriate
selection effects accounted for as in Figure~\ref{fig:NHdistrib}.
Finally, we note that although we do not see extremely large column
densities $\nh\gtrsim10^{24}\,{\rm cm^{-2}}$ in the distributions from
our simulations, our model does not rule out such values. It is
possible that very bright quasars in unusually massive galaxies or
quasars in higher-redshift, compact galaxies which we have not
simulated may, during peak accretion periods, reach such values in
their typical column densities. Moreover, as our model assumes
$\sim90\%$ of the mass of the densest gas is clumped into cold-phase
molecular clouds, a small fraction of sightlines will pass through
such clouds and encounter column densities similar to those shown for the
cold phase in Figure~\ref{fig:compareNHcalc}, as large as $\nh\sim10^{26}{\rm
cm^{-2}}$. This also allows a large concentration of mass in
sub-resolution obscuring structures, such as an obscuring toroid on
scales $\lesssim100\,$pc, although many of the phenomena such
structures are invoked to explain can be accounted for through our
model of time-dependent obscuration.

\section{Extension of Quasar Lifetimes to Other Masses\label{sec:lifetimes}}

\begin{figure}
    \centering
    %\plotone{f4.ps}
    \includegraphics[width=3.5in]{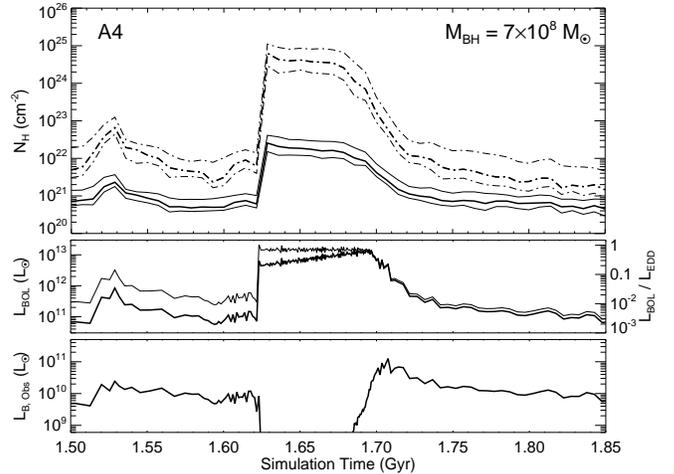}
    \caption{Upper panel: Thick contours plot the median column
    density \NH\ as a function of simulation time, with thin contours
    at 25\% and 75\% inclusion levels, for the A4 simulation ($V_{\rm
    vir}=226\,{\rm km\,s^{-1}}$).  Solid contours represent the density
    of the ``hot-phase'' ISM, dashed contours the total simulation
    density.  Middle panel: Bolometric luminosity of the
    black hole in the simulation, 
    $\Lbol=\dEdt$ (thick), and ratio of bolometric to
    Eddington luminosity, $l\equiv\Lbol/L_{Edd}$ (thin). Values are
    shown for each simulation timestep.  Lower panel: Observed B-band
    luminosity with \NHI\ calculated using the median ``hot-phase'' ISM density.
    \label{fig:NH.vs.time.A4}}
\end{figure}

\begin{figure}
    \centering
    %\plotone{f5.ps}
    \includegraphics[width=3.5in]{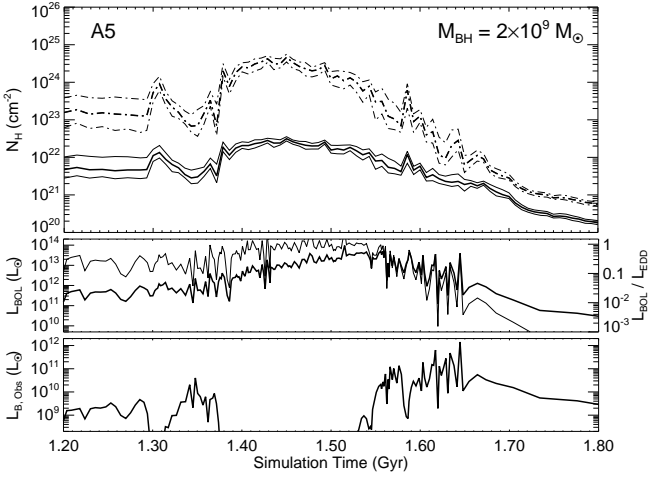}
    \caption{Same as Figure~\ref{fig:NH.vs.time.A4}, but for the A5
    ($V_{\rm vir}=320\,{\rm km\,s^{-1}}$) simulation.
    \label{fig:NH.vs.time.A5}}
\end{figure}

We generalize the results from \ctH\ and those above (which use
simulation A3) to a range of masses, using the simulations A1, A2, A4,
and A5. Recall, these simulations are identical except for increasing
galaxy mass given by the sequence in virial velocity, $V_{\rm vir}=80,
113, 160, 226, {\rm and}\ 320\,{\rm km\,s^{-1}}$ for A1-A5,
respectively. The evolution of these simulations would be 
simply related if not for the scale-dependent 
physics of cooling, star formation, and black
hole accretion and feedback. Figures~\ref{fig:NH.vs.time.A4} and
\ref{fig:NH.vs.time.A5} plot the bolometric luminosity of and column
densities to the supermassive black hole as a function of simulation
time, for the period shortly before and after the merger in
simulations A4 and A5, respectively. This is presented in the style of Figure 2 from
Hopkins et al. (2005a), which showed the same for A3. 
The median column densities and
25\%-75\% contours are indicated for the total average simulation density,
and the calculated hot-phase density, as described in \S2.2, %\,\ref{sec:NH}, 
along with the observed B-band luminosity using the
median hot-phase \NHI. The final masses of the black holes are also
shown, where $M_{\rm BH,\, final}\approx7\times10^{6}, 3\times10^{7},
3\times10^{8}, 7\times10^{8}, {\rm and}\ 2\times10^{9}$ for
simulations A1-A5, respectively. The evolution seen in simulation A4
is quite similar to that of A3, with a sharp rise at the time of the
merger to very large peak Eddington-limited luminosities, for a period
$\sim10^{8}\,{\rm yr}$, much of which is unobservable in the optical
owing to a corresponding rise in \NHI. At the final stages of strong
accretion, when gas is rapidly being consumed and expelled, the
observable luminosity finally rises to peak levels over a much shorter
time interval $\sim10^{7}\,{\rm yr}$.  In the A5 simulation, the
period of rapid accretion and large, Eddington-limited intrinsic
luminosities is significantly longer, $\sim3\times10^{8}\,{\rm yr}$,
and the peak luminosity is higher, as expected. However, as before,
most of this period is obscured by large column densities, and,
remarkably, once again only the final stages of intrinsically high
luminosity are visible, as gas is expelled and the accretion rate and
column density both fall. The same behavior is seen in the A1 and A2
cases, though with much lower peak bolometric luminosities
$\sim10^{10}$ and $10^{11}\,L_{\sun}$, respectively. The striking
similarity of this process across all our simulations suggests that it
is at least qualitatively insensitive to the details of the accretion
history and galaxy or black hole masses.

\begin{figure}
    \centering
    %\plotone{f6.ps}
    \includegraphics[width=3.5in]{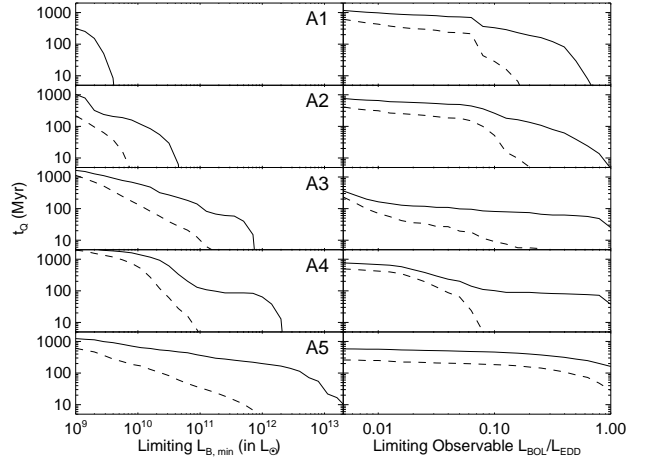}
    \caption{Left panels: Quasar lifetimes as a function of $\LBm$
    from the simulations (time with observed B-band luminosity $\LBo\geq \LBm$).
    Results from simulations A1-A5 are shown
    from top to bottom.  Solid lines indicate the lifetime if attenuation
    is ignored. The dashed lines give the median observed lifetime calculated with
    the metallicity-weighted, hot-phase neutral column density.  Right
    panels: Same as the left panels, but with \tq\ defined as a bolometric
    luminosity $\Lbol>l\,L_{Edd}$, where the observable lifetime is
    the time above a given ratio $l$ with B-band optical depth less than unity.
    \label{fig:lifetime.vs.L}}
\end{figure}

We quantify the resulting quasar lifetimes in
Figure~\ref{fig:lifetime.vs.L}, where we plot the quasar lifetime \tq\
as a function of the limiting B-band luminosity $\LBm$. The panels
show the lifetimes for simulations A1-A5, from top to bottom. In each
panel, the solid curve shows the intrinsic lifetime $\tQ^{\,i}$;
i.e. the total time the intrinsic $\LB\geq \LBm$.  The dashed curves
show the integrated time that the observed B-band luminosity meets
this criterion, using the median metallicity-weighted hot-phase
neutral column density of the simulation 
(case $C$ of Figure~\ref{fig:compareNHcalc}).  As demonstrated in
Figure~\ref{fig:compareNHcalc} and in \ctH, varying 
this definition of
the column density over realistic ranges 
can change the intrinsic lifetime by a factor of $\sim2-3$ 
at a given $\LBm$ . We also plot the lifetime as a function of $l$,
the ratio of the bolometric luminosity to the Eddington luminosity,
and find a similar trend. Here, we calculate the ``observed'' time
above a given ratio $l$ as the time above such a ratio with a
B-band optical depth less than unity. 
In all cases, the observed \tq\ is significantly
smaller than the intrinsic lifetime for all $\LBm>10^{9}\,L_{\sun}$,
and the ratio $\tQ/\tQ^{\,i}$ decreases with increasing
$\LBm$. However, both the intrinsic and observed lifetimes increase
systematically at a given $\LB>\Lcut{9}$ with increasing galaxy and
black hole mass, and the peak luminosity similarly increases, as
expected. In all cases corresponding to quasar-like luminosities, our
estimates of observed lifetimes of $\sim10^{7}\,{\rm yr}$ agree well
with observations, with much longer intrinsic lifetimes
$\gtrsim10^{8}\,{\rm yr}$.

A quasar accreting and radiating at the Eddington limit, with
luminosity increasing exponentially up to some peak luminosity
$L_{\rm max}$ after which accretion shuts off, will spend an integrated
time above a given luminosity $L$ given by
\begin{equation}
\tQ(L'>L)=t_{S}\,\ln{(L_{\rm max}/L)},
\end{equation}
where $t_{S}$ is the Salpeter time, $t_{S}=M_{\rm
BH}/\Mdot=4.5\times10^{7}\,$yr at the Eddington rate with our
efficiency $\epsilon_r=0.1$, and $L_{\rm max}$ is given by the Eddington
luminosity of the final black hole mass. This lifetime agrees well
with the plotted lifetimes for simulations A3, A4, and A5 at the
high-luminosity end. However, simulations A1 and A2 do not radiate at
their Eddington limits for a long period of time, and the slope of
this relation is too shallow in all cases, underpredicting the time
that the simulation spends at luminosities
$L\lesssim0.1\,L_{\rm max}$. This is because the black holes in the
simulations
spend significant time at lower Eddington ratios both going into and
coming out of the quasar stage and in an extended quiescent
phase. Moreover, the observed lifetimes do not at all correspond
to the lifetimes predicted by this simple model. Therefore, to simply
describe the intrinsic and observed lifetimes as a function of
limiting luminosity and final black hole mass (or peak luminosity
$L_{\rm max}$), we fit the calculated lifetimes \tq\ to truncated power
laws, from $L=\Lcut{9}$ to approximately $0.5\,L_{\rm max}$, or where
$\tQ\lesssim1\,$Myr, whichever occurs first.  Thus we fit to the form
\begin{equation}
\tQ(L'>L)=t_{9}\,(L/L_{9})^{\alpha}, 
\end{equation}
where $L_{9}\equiv\Lcut{9}$ and $t_{9}=\tQ(L'>L_{9})$. 

\begin{figure}
    \centering
    %\plotone{f7.ps}
    \includegraphics[width=3.5in]{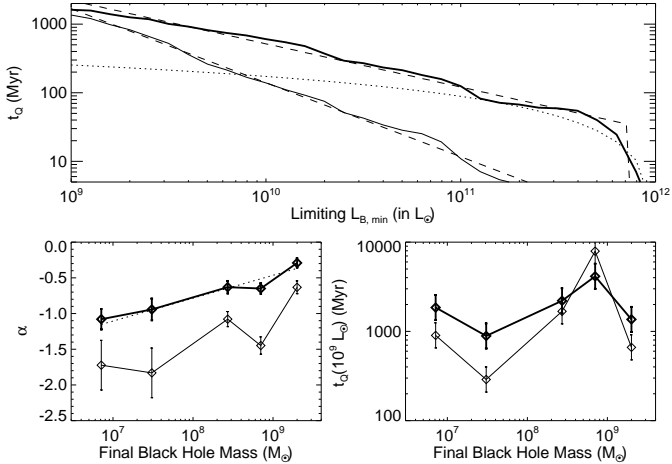}
    \caption{The results of fitting quasar lifetimes \tq\ to truncated
    power laws of the form $\tQ(L'>L)\propto L^{\alpha}$. Upper panel
    shows the intrinsic (thick solid line) and observed (thin solid
    line) lifetimes in the A3 simulation, with dashed lines indicating
    the corresponding best-fit truncated power laws. The dotted line
    shows the lifetime expected for Eddington-limited accretion up to
    the final black hole mass. The lower panels plot the power law
    slopes $\alpha$ and normalizations at $L=\Lcut{9}$ of the power
    law fits to the intrinsic (thick) and observed (thin) lifetimes
    from the series of simulations A1-A5, as a function of the final
    black hole mass in the simulation. The dotted line in the left panel gives a simple
    linear fit to $\alpha$ as a function of the logarithm of the final black hole mass.
    \label{fig:pwrlawfits}}
\end{figure}

Figure~\ref{fig:pwrlawfits} shows the resulting power-law slopes
$\alpha$ and normalizations $t_{9}$ of the lifetimes \tq\ as a
function of the final black hole mass (or equivalently, the peak
quasar luminosity given as the Eddington luminosity of the final black
hole mass).  The calculated, fitted, and Eddington-limit approximated
lifetimes are also shown for comparison for the A3 simulation. The
lifetimes are well-fitted by power laws in this range, and the
discrepancy between the power-law and Eddington-limit approximated
lifetimes at lower luminosities is clear.  The slope of both the
intrinsic and observed lifetimes becomes shallower at higher black
hole masses, as the lifetime increases for all luminosities and
extends to higher peak luminosities, but the slope of the observed
lifetime evolves much more rapidly than the slope of the intrinsic
lifetime. Over the range of our simulations, the intrinsic lifetime
slope can be approximated as
$\alpha=-0.79+0.32\,\log(M_{\rm BH}/10^{8}\,M_{\sun})$, or
$\alpha=-0.95+0.32\log(L_{\rm max}/10^{12}\,L_{\sun})$.  Although the
fluctuation in the normalization $t_{9}$ is large for observed
lifetimes, there does not appear to be a systematic trend with final
black hole mass, which is expected as by luminosities of this level or
lower, the lifetime is dominated by the quiescent-phase lifetime
throughout the duration of the simulation. This is especially clear in
the normalization of the intrinsic lifetimes, which also show no
systematic trend and are all within a factor of $\sim$ two of 2 Gyr, the
approximate total duration of the simulations.

In order to obtain a lower limit to quasar lifetimes at low
luminosities, we repeat this analysis, but ignore all times prior to the
intrinsically bright quasar phase (essentially the final merger).
This excludes the luminosity of the black hole as it grows from its
small seed mass, and ignores the periods of accretion with high gas
fractions early in the merger, giving a lower limit corresponding to
an already-large black hole suddenly ``turning on'' in a quasar
phase. As expected, the Eddington-limited phase of the lifetime is
more dominant in this case. However, the pure Eddington-limited model
still gives lifetimes too short by an order of magnitude at low
luminosities. Fitting to power laws we find very similar relations to
those shown in Figure~\ref{fig:pwrlawfits}, with slightly shallower
slopes, $\alpha=-0.61+0.30\,\log(M_{\rm BH}/10^{8}\,M_{\sun})$ for the
intrinsic lifetime, and normalization $t_{9}\sim1\,$Gyr. We further
fit one additional functional form to the intrinsic lifetimes, in
which the lifetime follows the Eddington-limited shape for large
luminosities, down to some luminosity $L_{break}$, and below this
luminosity follows a power-law. This allows a continuous
description of the lifetime, very accurate at both low and near-peak
luminosities.  For these fits, we find similar power-law slopes
$\alpha=-0.81+0.32\,\log(M_{\rm BH}/10^{8}\,M_{\sun})$, with
$L_{break}\approx0.025\,L_{bol,\,max}\approx0.14\,L_{B,\,max}$ nearly
constant.  We also fit the lifetimes as a function of frequency for
the representative visual wavelengths shown in
Figure~\ref{fig:lifetimeVSfreq}. Although, as expected from the
figure, the observed lifetime slopes become slightly shallower and the
intrinsic lifetime slopes become slightly steeper as the quasar is
observed at redder wavelengths, the change with wavelength across this
range is very small compared to the change with black hole
mass. Moreover, the change in normalization $t_{9}$ across visual
wavelengths is consistent with zero. We therefore find that quasar lifetimes
can be well approximated by simple power laws, 
where the power law slope depends on the final black hole mass or peak 
luminosity of a system, and find significantly longer 
lifetimes at low luminosities than predicted by simple models of 
accretion at a constant Eddington ratio. 
Furthermore, attempts to 
de-convolve intrinsic quasar properties and observations as well as
semi-analytical attempts to model observable quasar luminosity 
functions and statistics require an accurate model of quasar 
lifetimes as a function of luminosity and host system properties. The 
fits we describe have the advantage of a simple, analytical 
form which can be applied to a continuum of luminosities and 
intrinsic quasar properties, while addressing the shortcomings of 
simpler lifetime models below peak system luminosities.

\section{Discussion \&\ Conclusions\label{sec:discussion}}

In this paper, we have studied implications of the conclusion of \ctH\
that incorporating the effects of obscuration in a galaxy merger
simulation gives observed quasar lifetimes of $\sim10^{7}\,{\rm yr}$.
We extended our analysis to a series of five simulations with virial
velocities $V_{\rm vir}=80-320\,{\rm km\,s^{-1}}$, producing a range
of final black hole masses from $\sim7\times10^{6} -
2\times10^{9}\,{\rm M_{\sun}}$, and find qualitatively similar results
in all cases and for different determinations of the obscuring column
density.  The good agreement with observations and the significant
difference between observed and much longer intrinsic lifetimes is
found in all our simulations.  The qualitative evolution of quasars in
this model is robust, with the result reproduced across a range of
galaxy masses, and using different methods for calculating column
densities that can give factors greater than an order of magnitude
different attenuations along any given line of sight at a particular
time. We therefore expect that quasars will have extended intrinsic
lifetimes, much of which are obscured in visual wavelengths.  The
processes which fuel periods of quasar activity by channeling
significant quantities of gas into the central regions of a galaxy
will produce this extended intrinsic lifetime of $\gtrsim10^{8}\,$yr.
However, for a significant fraction of the intrinsic lifetime for
quasar activity, this same process will produce large column densities
which heavily obscure the quasar and attenuate it well below
observable limits in the B-band and other visual
wavelengths. Eventually, feedback from accretion energy will remove
surrounding gas, creating a window in which the quasar is optically
observable before accretion rates drop below those needed to maintain
quasar luminosities.

Of course, not all aspects of AGN activity are related
to mergers.  For example, some low redshift quasars (e.g. Bahcall et
al. 1996) and many Seyferts appear to reside in relatively ordinary
galaxies. Gas expelled by quasar feedback may cool and 
relax to the central regions of the galaxy, and passive stellar evolution 
can produce a large fuel supply in winds. 
In our picture, it is most natural to interpret occurrences of AGN
activity that are not caused by mergers as arising from
normal galaxies being re-activated as gas is sporadically accreted by a
black hole created earlier in a bright quasar phase.  
The principle requirement of this modeling 
is that such activity should not contribute a
large fraction of the black hole mass, to avoid spoiling tight
correlations between the black hole and host galaxy properties or 
overproducing the present-day density of very massive black holes. 
Radiative heating from the final stages of quasar 
activity may suppress much of this activity (much of the 
remaining gas is heated to very high temperatures; e.g.\ Cox et al.\ 2005), and  
the hydrodynamical modeling of \citet{CO97,CO01,Sazonov05} 
demonstrates that residual accretion from stellar feedback and 
``cooling flows'' is dramatically suppressed by even relatively small AGN 
feedback once the black hole has reached a critical mass determined by 
the $M_{\rm BH}$-$\sigma$ relation. 

Further, demographic arguments such as that of \citet{Soltan82}, 
as well as more detailed analysis by e.g., \citet{YT02}
implies that most of the present black hole mass density 
is accumulated in bright quasar phases, for which our modeling 
should be most applicable. This remains true in our analysis, 
and we demonstrate in Hopkins et al.\ (2005b,c) that our modeling 
can account for the entire quasar luminosity function 
in different frequencies, although we defer a detailed prediction 
of the resulting supermassive black hole density and mass function 
to a future paper. It is important to note, however, that 
given our prediction of an obscured 
accretion phase with a duration up to $\sim10$ times that of the 
optically observable phase, a naive application of the 
\citet{Soltan82} or other demographic arguments would imply 
a similar increase in the black hole mass density accumulated in 
obscured quasar phases. But this neglects both the fact that black holes 
are growing during this phase and the luminosity dependence of 
the quasar lifetime. A more detailed calculation shows that a 
large fraction ($\gtrsim80\%$ in the brightest quasars) of the quasar 
mass is accumulated in the optically observable stages of 
black hole growth when the 
final $e$-folding of exponential black hole growth occurs as the black hole 
begins to drive strong feedback and soon shuts down accretion. Similarly, 
most of the total black hole radiated energy 
is emitted in this stage, with a considerable fraction ($\sim0.3-0.6$) 
observable. Thus, we expect that corrections to the estimates of 
black hole growth from demographic arguments based on observations 
of bright, relatively unobscured quasars will in fact only
be of order unity from our modeling, and thus should 
not conflict with theoretical limits. It is still an interesting and 
important question, however, to determine from our modeling the implied 
relic AGN density and constraints from the combination of observed
luminosity functions and black hole mass functions. 

We do not expect other considerations, such as the orbital parameters
of the merger or the collimation of black hole feedback, to
substantially change our results. Different orbital separations,
orientations, and energies will change the time of interaction, but
ultimately our model for quasar evolution depends only on the merger
eventually depositing sufficient gas into the central regions of the
remnant to fuel rapid black hole growth. After such a process begins,
the evolution of the black hole should be determined by the
self-regulated mechanisms which terminate black hole growth, as
described in \citet{DSH05}. This will eventually expel gas, rendering
the quasar observable and shutting down the accretion phase at the
final black hole mass given by the $M_{\rm BH}$-$\sigma$ relation. The
phase of rapid black hole growth and the expulsion of gas in the final
stages of strong accretion occur during the end-stages of the merger and
in the very center of the merging cores, when and where the structure
of the original orbits should be least important.  This view is supported
by additional simulations not presented here but shown in e.g.
Fig. 16 of \citet{SDH05b}, demonstrating that the final black hole
mass is relatively insensitive to orbital parameters.

The exact mechanism of black hole feedback should similarly not
dramatically change this picture, so long as a period of rapid black
hole growth occurs, requiring large densities (which naturally
generate obscuring column densities), and ultimately terminates via
some self-regulated mechanism which expels gas from central regions,
rendering the quasar optically observable in the final stages of its
life. In particular, collimation of the black hole feedback may
significantly increase the energy or momentum input on-axis, ``blowing
out'' material along this axis earlier in the quasar
lifetime. However, since the black hole is growing exponentially
during peak accretion, this should occur at most on the order of a
couple of Salpeter times (during which the black hole luminosity grows
by an order of magnitude) before it would in the isotropic case,
changing the intrinsic lifetime by only a fraction of its total.  The
unobscured stage of the quasar lifetime may then begin slightly
earlier along on-axis sightlines, but should be of comparable duration
as a similar mechanism removes the gas from central regions. In
addition, in order for significant black hole growth to cease at the
same critical mass as in the isotropic case (determined by the $M_{\rm
BH}$-$\sigma$ relation), the high-accretion rate phase must end over
the same timescale $\sim t_{S}$ as the ``lead'' with which collimated,
on-axis blowout precedes isotropic blowout.  The necessity that
self-regulation eventually terminates accretion implies that
surrounding densities, even off-axis, must decrease, resulting in at
most a broadening of the column density distribution as typical column
densities off-axis could be somewhat larger than column densities
on-axis.

For much of this regime, we find that the source should be observable
in hard X-ray frequencies, as the attenuation is weaker and column
densities generally lie below the Compton thick limit. This
distinction naturally produces a substantial population of obscured
quasars, as a standard phase in the evolution of quasars over their
lifetimes.  Although we defer a detailed calculation of
the quasar luminosity function as a function of redshift and in
different observed frequencies (Hopkins et al. 2005b), 
we note that such a population can
naturally account for both the cosmic X-ray background spectrum and
discrepancies between optical and X-ray luminosity functions at
various redshifts (Hopkins et al. 2005c). 
Moreover, optical and X-ray observational
samples will be affected rather differently by selection effects and
magnitude limits, especially for quasars near their peak
luminosities. Therefore, the differences in the evolution of optical
and X-ray selected samples may be accounted for through the dependence
of obscuration on intrinsic luminosity and host galaxy properties.
Attenuation is also weaker in the infrared, and the large fraction of
optical and UV energy absorbed during the obscured phase may be
reprocessed by dust, appearing as thermal radiation in the IR. This
may render the observable lifetime larger than the intrinsic
lifetime at long wavelengths, an effect important for calculating the
IR background and estimating source populations. Similarly, the total
obscuration and obscuration as a function of intrinsic luminosity are
important for the evolution of luminous and ultraluminous infrared
galaxies, and for estimating the relative energy contributions in the
IR of starbursts and active galactic nuclei (AGN).

The typical column densities in our simulations correspond well to
observed column densities of optically selected quasars, once the
appropriate observed magnitude limit has been imposed. Thus, our model
allows us to reproduce the distribution of optically obscured sources
above this magnitude limit naturally, as an evolutionary effect of the
mechanisms which fuel and regulate quasar growth, without invoking
particular distributions of source properties or geometric patterns of
obscuration. Moreover, the column densities we calculate in our
simulations allow us to predict that the distribution of \NH\ values
should become broader as the minimum observable luminosity is
decreased, as both faint, quiescent phases with low
($\lesssim10^{20}\,{\rm cm^{-2}}$) column densities and bright,
obscured phases with high ($\gtrsim10^{22}\,{\rm cm^{-2}}$) column
densities become observable. This effect is seen in the broad
distribution of \NH\ values in X-ray samples, which are much less
affected by attenuation, but a more detailed analysis including
modeling the distribution of quasar properties is needed to reproduce
both the optical and X-ray \NH\ distributions more accurately.

We find that the peak luminosities and lifetime above any given
luminosity increase systematically with galaxy mass and final black
hole mass, although the distinction between observed and intrinsic
lifetimes remains significant in all cases. Moreover, the ratio of
observed to intrinsic lifetimes decreases in all cases with increasing
minimum luminosity. We also find that intrinsic lifetimes at
luminosities $\lesssim0.1$ times the peak luminosity are poorly fit by
assuming quasars always accrete near the Eddington limit, but rather
that lifetimes are well-fitted by power laws with a steeper
slope. This is a result of sub-Eddington accretion rates both before
and after the peak accretion phase, even at luminosities significantly
above the quiescent steady-state small accretion rates seen at late
times in the simulations. This suggests that many quasars seen at low
luminosities may be quasars with a large peak luminosity in a
significantly sub-Eddington phase, although the fraction of quasars
observed at high Eddington ratios becomes large with increasing
luminosity.

Additionally, these lifetimes imply that, even at high luminosities
where growth may be Eddington-limited, quasars spend a significant
fraction of their lives with intrinsic luminosities well below their
peak luminosities. Therefore, any observed luminosity function is the
convolution of the distribution of quasars with a given peak
luminosity (determined by the final black hole mass and thus the
merging galaxy properties) and a non-trivial light curve
(Hopkins et al. 2005b). It is clear
from these calculations that any attempt to theoretically model even
the intrinsic luminosity functions of quasars must take into account
the functional dependence of the light curve on luminosity and
time.  In order to apply these models to observed luminosity
functions, the dependence of observed luminosity on intrinsic
luminosity and the quasar evolution, as well as the observed
frequency, must be considered (Hopkins et al. 2005b,c).

We find that the ratio of observable to intrinsic lifetimes is a
strongly decreasing function of the limiting luminosity of
observations.  Modeling this effect is necessary to estimate intrinsic
quasar lifetimes from observations, as well as for using theoretically
motivated accretion models to predict the quasar luminosity function
and space density of present-day supermassive black holes. Further,
the effects we describe can account for the presence of an obscured
population of quasars which are missed by optical, UV, or soft X-ray
surveys but may contribute significantly to the cosmic X-ray
background.  Observations of the cosmic X-ray background and
comparison of the optical and hard X-ray quasar luminosity functions
that indicate the existence of a large obscured population of quasars
\citep[and references therein]{BH05}, is explained in our picture,
because it predicts different observed lifetimes and populations at
different frequencies.  The scenario we describe also implies that the
reprocessing of quasar radiation by dust in surrounding regions can
account for observations of luminous and ultraluminous infrared
galaxies with merger activity and obscured AGN.

Together with the modeling presented by Di Matteo et al. (2005),
Springel et al. (2005a,b), Springel \& Hernquist (2005),
Hopkins et al. (2005a,b,c), and Robertson
et al. (2005), the results
described here motivate the following picture for galaxy formation and
evolution.
%, illustrated schematically as a ``cosmic cycle'' in
%Figure~\ref{fig:cosmiccycle}.  
Through the hierarchical growth of
structure in a cold dark matter universe (White \& Rees 1978), mergers
between galaxies occur on a regular basis.  Those involving gas-rich
progenitors, which would be increasingly more common towards higher
redshifts, produce inflows of gas through gravitational torques
\citep{BH91,BH96}, causing starbursts \citep{MH94,MH96} like those
associated with luminous infrared galaxies (e.g. Sanders \& Mirabel
1996).  The high gas densities triggering these starbursts fuel rapid
black hole growth.  For most of the period over which black hole
growth occurs, optical quasar activity would be buried, but X-rays
from the black holes explain the presence of non-thermal point sources
in e.g. NGC 6240 \citep{Komossa03}, and reprocessing of most of the
black hole energy by surrounding gas and dust can, in principle,
account for the spectral energy distributions of ``warm'' ULIRGs
(Sanders et al. 1988c) and recent observations of a correlation between 
quasar obscuration and far-infrared host luminosity 
\citep[e.g.,][]{Page04,Stevens05}.  
As the black hole mass and radiative output
increase, a critical point is reached where feedback energy starts to
expel the gas fueling accretion.  For a relatively brief period of
time, the galaxy would be seen as an optical quasar with a B-band
luminosity and lifetime characteristic of observed quasars.  This
phase of evolution is brief ($\sim 10^7$ yr), owing to the explosive
nature of the final stages of black hole growth as the gas responds
dramatically to the feedback energy from the exponentially evolving
black hole.  This AGN feedback terminates further black hole growth,
leaving a remnant that resembles an ordinary galaxy containing a dead
quasar and satisfying the $M_{\rm BH}$-$\sigma$ relation.  
Modeling both the dependence of quasar lifetime on luminosity, and the 
complex, time-dependent evolution of quasar obscuration is thus crucial to 
any observational or theoretical understanding of quasars
and quasar host galaxy evolution.

\acknowledgments
This work was supported in part by NSF grants ACI
96-19019, AST 00-71019, AST 02-06299, and AST 03-07690, and NASA ATP
grants NAG5-12140, NAG5-13292, and NAG5-13381.
The simulations
were performed at the Center for Parallel Astrophysical Computing at the 
Harvard-Smithsonian Center for Astrophysics.

%\begin{figure}
%    \centering
%    \plotone{f8.ps}
%    %\includegraphics[width=3.2in]{f8.ps}
%    \caption{Schematic representation of our ``cosmic cycle'' for
%    galaxy formation and evolution regulated by black hole
%    growth in mergers.
%    \label{fig:cosmiccycle}}
%\end{figure}

\end{document}